\documentclass[twocolumn,showpacs,preprintnumbers,amsmath,amssymb]{revtex4}

\usepackage{graphicx}
\usepackage{dcolumn}
\usepackage{bm}

\begin{document}

\title{Diffusion entropy analysis on the scaling behavior of financial markets}
\author{Shi-Min Cai$^{1}$}
\author{Pei-Ling Zhou$^{1}$}
\author{Hui-Jie Yang$^{2}$}
\author{Chun-Xia Yang$^{1}$}
\author{Bing-Hong Wang$^{2}$}
\author{Tao Zhou$^{1,2}$}
\email{zhutou@ustc.edu}

\affiliation{$^{1}$Department of Electronic Science and
Technology, University of Science and Technology of China, Hefei
Anhui, 230026, PR China\\ $^{2}$Department of Modern Physics,
University of Science and Technology of China, Hefei Anhui,
230026, PR China
\\
}

\date{\today}

\begin{abstract}
In this paper the diffusion entropy technique is applied to
investigate the scaling behavior of financial markets. The scaling
behaviors of four representative stock markets, Dow Jones
Industrial Average, Standard\&Poor 500, Heng Seng Index, and Shang
Hai Stock Synthetic Index, are almost the same; with the
scale-invariance exponents all in the interval $[0.92, 0.95]$.
These results provide a strong evidence of the existence of
long-rang correlation in financial time series, thus several
variance-based methods are restricted for detecting the
scale-invariance properties of financial markets. In addition, a
parsimonious percolation model for stock markets is proposed, of
which the scaling behavior agrees with the real-life markets well.
\end{abstract}

\pacs{89.90.+n,05.10.-a,05.45.Tp,64.60.Cn,87.10.+e}

\maketitle

\section{Introduction}
Analysis of financial time series attracts special attentions from
diverse research fields for several decades. It can not only
reveal the intrinsic dynamical properties of the corresponding
financial markets but also provide us a clear scenario to
construct dynamical models. Traditional theories are constructed
based upon some basic hypothesis, to cite examples, the stochastic
processes in the markets and the homogenous property of the
markets, etc. The unexpected so-called rare events are explained
simply as the results due to accidents or external
triggers\cite{Sornette2000}. The advancements in nonlinear theory
lead a complete revolutionary in our ideas about financial
markets. Instead of the deduced Gaussian distribution, empirical
investigations in recent years indicate that the price return
distribution of the financial time series generally obeys the
centered L\'{e}vy distribution and displays fat-tail property, and
the financial time series exhibits the scale-invariance behavior
\cite{Mandelbrot1963,Mantegna1995,Mantegna1997,Lo1999J,Wang2001,Stanley2001}.
The nonlinear theory based analysis and dynamical models for the
financial markets are the essential problems at present time.

One of the important features of the financial time series is the
scale-invariance property, which can highlight the dynamical
mechanics for the corresponding markets. Consider a complex system
containing a large amount of particles. The scale-invariance
property in the diffusion process of this system can be described
mathematically with the probability distribution function as
\begin{equation}
P(x,t)=\frac{1}{t^\delta}F(\frac{x}{t^\delta}),
\end{equation}
where $x$ is the displacements of the particles in the complex
system and $\delta$ the scale$-$invariance exponent. The
theoretical foundation of this property is the Central Limit
Theorem and the Generalized Central Limit Theorem
\cite{Ma1985,Gnedenko1954}. For $\delta=0.5$, the diffusion
process is the standard diffusion and $F(y)$ is the Gaussian
function. And $\delta\neq0.5$ exhibits the deviation of the
dynamical process from the normal stochastic one. For a financial
time series, the delay$-$register vectors, denoted with
$\{y_{k},y_{k+1},\cdot\cdot\cdot,y_{k+m-1}|k=1,2,3,\cdot\cdot\cdot,N-m+1\}$,
can be regarded as the trajectories of $N-m+1$ particles during
the period of $0$ to $m$. By this way we can map a time series to
a diffusion process, called overlapping diffusion process in this
paper. An alternative solution is to separate the considered time
series into many non-overlapping segments and regard these
segments as the trajectories.

In literature, several variance-based methods are proposed to
detect the scale-invariance properties, such as the probability
moment method \cite{Paladin1987}, the fluctuation approach and the
de-trended fluctuation approach \cite{Peng1994}, etc. But these
variance-based methods have two basic shortcomings. One is that
the scale-invariance property can be detected but the value of the
exponent cannot be obtained correctly. The other is that for some
processes, like the L\'{e}vy flight, the variance tends to
infinite and these methods are unavailabel at all. Although the
infinite can not be reached due to the finite records of empirical
data, clearly we can not obtain correct information about the
dynamics under these conditions. Hence, in this paper we suggest
the using of diffusion entropy (DE) technique to detect the
scaling behavior of financial markets.

\section{Diffusion Entropy Technique and Data Analysis}
To overcome the above shortcomings in the variance-based methods,
the authors in reference\cite{Scafetta2002} designed the diffusion
entropy analysis (DEA). To keep our description as self-contained
as possible, we review the DEA method briefly.

Filter out the trends in the original time series. Adopting the
traditional assumption generally used in the research filed of
engineering, that a discrete time series variable consists of a
slowly varying part and a fluctuation
part\cite{Ignaccolo20041,Ignaccolo20042}, the index of a stock
market reads,
\begin{equation}
\xi_j=S_j+\zeta_j,j=1,2,\cdots,N,
\end{equation}
where $\zeta_j$ is the fluctuation with zero mean and fixed
variance.

In the signal processing, the slow and regular variation $S_j$ is
usually called signal, which contains the useful information. And
the rapid erratic fluctuation $\zeta_j$ is called the noise, which
is regarded as perturbations containing only trivial information.
In the DEA method, however, the scale$-$invariance will be
detected from this fluctuation part. The $"$step
smoothing$"$\cite{Ignaccolo20041} procedure is employed to
estimate the $S_j$ part in our calculations, that is, regard the
average of the segments as the trends, respectively. The final
time series is regarded as a steady series, whose overlapping
diffusion process reads,
\begin{equation}
x_k(t)=\sum^{k+t}_{j=k}\zeta_j,k=1,2,\cdots,N-t+1.
\end{equation}
Consequently, the Shannon entropy can be defined as,
\begin{equation}
S(t)=-\int^{+\infty}_{-\infty}P(x,t)\log_{10}[P(x,t)]dx.
\end{equation}
 A simple algebraic leads to,
\begin{equation}
S(t)=A+\delta\log_{10}(t),
\end{equation}
where
\begin{equation}
A=-\int^{+\infty}_{-\infty}F(y)\log_{10}[F(y)]dx,y=\frac{x}{t^\delta}.
\end{equation}
The DEA method has been used to deal with many time series in
different research fields, such as the solar induced atmosphere
temperature series\cite{Grigolini2002}, the intermittency time
series in fluid turbulence\cite{Bellazzini2003}, the spectra of
complex networks\cite{Yang2004}, and the output spike trains of
neurons\cite{Yang2005}.

In order to truly uncover the scaling behavior of financial
market, we study both the domestic and oversea stock markets. The
domestic consist of Shang Hai Stock Synthetic Index (SS) with
length $N=5412$, and Hang Seng Index (HSI) with length $N=191076$,
while Dow Jones Industrial Average (DJIA) with length $N=29627$
and Standard\&Poor 500 (S\&P500) with length $N=5695$ are
constituted of the oversea. In Fig. 1, the DE technique is applied
to analysis these real financial time series and we demonstrate
the scaling behavior of different Indices are almost same,with the
scaling exponents all in the interval $[0.92, 0.95]$; as showing
below: $\delta_{ss}=0.923\pm0.004$, $\delta_{HSI}=0.944\pm0.004$,
$\delta_{DJIA}=0.948\pm0.005$, and
$\delta_{S\&P500}=0.950\pm0.005$. These results provide a strong
evidence of the existence of long-rang correlation in financial
time series, thus several variance-based methods are restricted
for detecting the scale-invariance properties of financial
markets.

\begin{figure}
\scalebox{0.8}[0.8]{\includegraphics{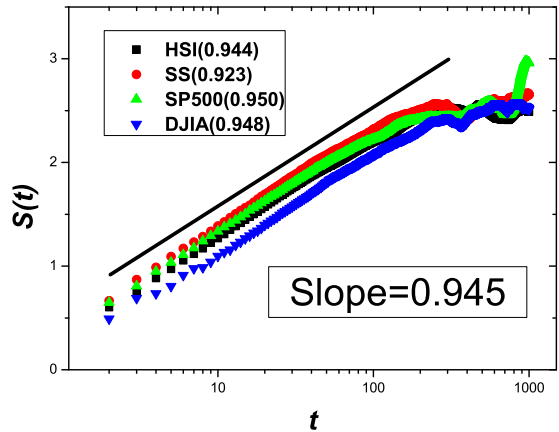}}
\caption{\label{fig:epsart} (Color online) DE results for four
time series from different stock markets. The squares, circles,
upward triangles and downward triangles denote the results for
HIS, SS, S\&P500 and DJIA, respectively. The corresponding scaling
exponents are $\delta_{HSI}=0.944\pm0.004$,
$\delta_{SS}=0.923\pm0.004$, $\delta_{S\&P500}=0.950\pm0.005$ and
$\delta_{DJIA}=0.948\pm0.005$, The solid line, whose slope is
$\delta_{fit}=0.945$, is presented as reference. The occurrence of
saturation regime caused by the $"$Step smoothing$"$ is $t=365$.
Consequently, the fitting interval should be in the range of
$t<\frac{T}{3}$\cite{Ignaccolo20041}. In the fitting procedure is
implemented in the interval of $t\in[1,100]$}
\end{figure}

Herein it must be noted that the occurrence of saturation regime
caused by the $"$Step smoothing$"$ is $t=365$. Consequently, the
fitting interval must be only a limited range of
times,$t<\frac{T}{3}$\cite{Ignaccolo20041}, which we estimate to
be $t\in[1,100]$.

\section{The Model}
There are many modelling methods to explain the origins of the
observed nonlinear scaling behavior of market price as emerging
from simple behavioral rules of a large number of heterogeneous
market participants, such as behavior-mind
models\cite{Thaler1993,Lo1999}, dynamic-games
models\cite{Friedman1991,Zhou2005}, cellular-automata
models\cite{Wei2003,Zhou2004}, multi-agent
models\cite{Lettau1997,Chen2001}, and so on. Here in this paper,
we consider a stock market model based on percolation
theory\cite{Grimmett1989}. Cont and Bouchaud\cite{Cont2000}
successfully applied percolation theory to modelling the financial
market(CB model), which is one of the simplest models able to
account for the main stylized fact of financial markets, e.g. fat
tails of the histogram of log-returns. Up to now, the percolation
theory is widely used in modelling stock
markets\cite{Stauffer2000,Castiglione2001,Makowiec2004,Wang2005}.

Based on percolation theory, our model incorporates the following
components different from the original CB model: (1) The cluster,
defined as a set of interconnected investors, grows in a
self-organized process. (2) The effect of $"$herd behavior$"$ on
the trade-volume is magnified step by step during the cluster's
self-organized accumulating process rather than instantaneously
formed like EZ model and its
variety\cite{Eguiluz2000,Zhengdafang2002,zhengbo2004,Xie2005}. (3)
Some encountering smaller clusters will form a bigger cluster
through cooperating or one defeating the rivals. (4) An infinite
cluster maybe exist without the need to tune $p$ to $p_c$ and its
trade activity influence price fluctuation. (5)The number of
investors participating in trading will vary dynamically. Now
let's see the model.
\subsection{Dynamic of investor groups}
Initially, $M$ investors randomly take up the sites of an $L\ast
L$ lattice. Then for each cluster, a strategy is given: buying,
selling, or sleeping, which are denoted by 1, -1 and 0
respectively. In reality, the circle of professionals and
colleagues to whom a trader is typically connected evolves as a
function of time: in some cases, traders are following strong
herding behavior and the effective connectivity parameter $p$ is
high; in other cases, investors are more individualistic and
smaller values of $p$ seem more reasonable. In order to take the
complex dynamics of interactions between traders into account, we
assume that it undergoes the following evolution repeatedly:

(1)Growth: most of investors would like to imitate the strategies
which have been adopted by many others, which induces $"$herd
behavior$"$ occurring. In this sense the herd behavior is
amplified. Specially, the affection of the herd behavior will be
magnified gradually with the increase of the number of investors
adopting this strategy, i.e., with the growth of the clusters.
During cluster's growth, a number of new investors will be
attracted by it and become its members. In other words, every
cluster will absorb new investors with the probability
\begin{equation}
P_d(\tau)=P_d(\tau-1)+k(N_T-N_T(\tau-1)),
\end{equation}
where $k$ is a kinetic coefficient controlling the growth speed
and $N_T$ is a threshold parameter (It has been validated that the
value of the parameters $k$ and $N_T$ could be any value. These
two parameters have no effects on the self-organization process of
the clusters\cite{Cavalcante2002}). $N_T(\tau-1)$ is the number of
the agents along its boundary, defined as a set made up of agents
which belong to a cluster and at least border on a site which
isn't part of this cluster, at the last time step $\tau-1$. The
new participating investor will take up the empty sites around the
old clusters and imitate the same strategy as that of it. The
probability $P_d$ is obviously limited to the range [0,1] so that
we have to impose $P_d=0$ and $P_d=1$, if the recurrence
relationship Equ.7 gives values for $P_d<0$ or $P_d>1$.

(2) New cluster's birth: some investors will randomly and
independently enter the market with the probability $P_n$. These
investors don't belong to an arbitrary existing cluster and will
take up the empty sites.

(3) Cooperation: encountering clusters will operate cooperation
and confliction between them. When their strategies are same, they
are thought to cluster together to form a new group of influence.
Or there would be confliction between them. The consequence of
confliction is that losers would be annexed by the winner and that
a new and bigger cluster whose strategy inherent the winner's will
be formed. The probability of cooperation or confliction is as
follow, i.e., some a cluster will cooperate with or defeat others
with the probability
\begin{equation}
P_m(k)\sim |s_\tau^k|,
\end{equation}
where $|s_{\tau}^{k}|$ is the size of k-th cluster at time $\tau$.

(4) Metabolism: in reality, no matter how huge has the size of a
group ever been it would collapse due to different influences such
as government decision on war and peace. Some new clusters will
come into the world in the wake of aging clusters' death. The
probability with which an aging clusters will die
is:\vspace*{-5pt}
\begin{equation}
P_o=\frac{x+y}{2L},
\end{equation}
where $x$ and $y$ is the width of this cluster occurring on the
lattice in the $x$ and $y$ direction. Equ.(4) indicates that the
probability with which a cluster disbands would increase with the
cluster growth. Once a spanning cluster exists, it would surely
die. When a cluster disbands, all its members would leave the
market and the sites where the death cluster ever occupied will be
taken up by new investors with the probability $P_n$. Such
occupied sites form a few new clusters. Every new cluster would be
given a strategy randomly.

\begin{figure}
\scalebox{0.8}[0.8]{\includegraphics{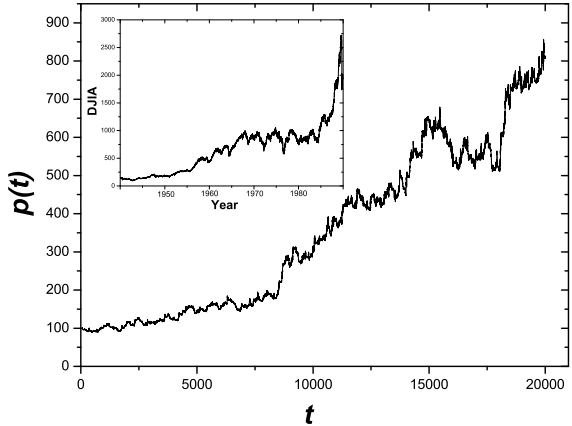}}
\caption{\label{fig:epsart} (Color online) Time series of the
typical evolution of the stock price in the interval . The insect
is the Dow Jones Industrial Average (DJIA) from 01-02-1940 to
12-31-1987.}
\end{figure}

Although each cluster could trade with others at every trading
step, the evolution frequency of the network topology should not
be so often. Thus, we assume that the network structure of the
market composed by investor groups would evolve every $N$ trading
steps. With the evolution of this artificial stock market, the
number of investors participating in trading isn't constant. The
network will take on different structure; the affection of the
herd behavior on the trade-volume is gradually magnified. Without
any artificial adjustment of the connectivity probability $p$ to
$p_c$, spanning cluster may exist, whose activity would influence
the price fluctuation.
\subsection{Trading rules}
Each cluster trades with probability $a$ $($called activity$)$; if
it trades, it gives equal probability to a buying or selling with
demand proportional to the cluster size. The excess demand is then
the difference between the sum of all buying orders and selling
orders received within one time interval. The price changes from
one time step to the next by an amount proportional to the excess
demand. To explain the $"$volatility$"$, Stauffer introduces the
feedback mechanism between the difference of the $``$supply and
demand$"$ and activity of the investors\cite{Stauffer2000}.
Whereas in our model, the difference of the ``supply and demand"
not only affects the activity probability but also the probability
with which the active clusters choose to buy or sell. The
probability $a$ evolves following the Equ.(5):
\begin{equation}
a(t)=a(t-1)+lr(t-1)+\alpha,
\end{equation}
where $r$ is the difference between the demand and supply, $l$
denotes the sensitivity of $a$ to $r$ and $\alpha$ measures the
degree of impact of external information on the activity. Each
active cluster choose to buy or sell with probabilities
$\frac{1}{2}a(t)(1-p_s(t))$ and $\frac{1}{2}a(t)p_s(t)$
respectively. For $r>0$, $p_s(t)=0.5+d_1r(t-1)$, while for $r<0$,
$p_s(t)=0.5+d_2r(t-1)$. According to Kahneman and
Tversky\cite{Kahneman1979}, it is asymmetry that agents make their
decisions when they are in face of gain or loss. When referring to
gain, most of the agents are risk adverse. On the contrary, they
are risk preference. These determine the parameters $d_1$ and
$d_2$, representing the sensitivity of the agent's mentality to
the price fluctuations and differing from each other. In our model
we assume $d_2=2d_1$. The difference between the demand and supply
is:\vspace*{-5pt}
\begin{equation}
r(t)=\sum_{j=1}^m\texttt{sign}(s_t^j)(|s_t^j|)^\gamma
\end{equation}
where $m$ is the total number of clusters occurring on the market.
$\gamma$ measures the degree of impact of each cluster's
trade-volume on the price, $0<\gamma<1$ allowing for a nonlinear
dependence of the change of $($the logarithm of the$)$ price as a
function of the difference between supply and
demand\cite{Farmer2002}. So the evolution of the price
is:\vspace*{-5pt}
\begin{equation}
P_r(t)=P_r(t-1)\exp(\lambda\emph{r}(t))
\end{equation}

\begin{figure}
\scalebox{0.8}[0.8]{\includegraphics{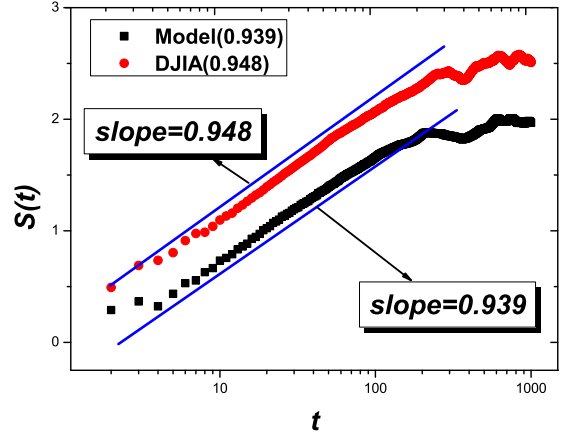}}
\caption{\label{fig:epsart} (Color online) The DE result, the
black squares denote the time series generated by the present
model, with the scaling $\delta_{model}=0.939$, which is almost
the same as that of DJIA denoted by red circles.}
\end{figure}

\subsection{Simulation}
When proper initial condition and parameters have been chosen, the
artificial stock market can generate its stock price. In Fig. 2 we
present a typical simulation result about price time series
generated by our model, which is rather similar to the real-life
index (inset). The parameters used here are $a(0)=0.09$,
$r(0)=0.09$, $p_r(0)=1$, $P_d(0)=0.4$, $k=0.0001$, $N_T=50$,
$l=\lambda=\frac{1}{L^{2}}$, $L=100$, $d_1=0.00005$,
$\gamma=0.78$, $P_n=0.6$, $M=100$, $N=50$.

By numerical studies, we have demonstrated that this model
exhibits the stylish facts with the price returns distribution is
a L\'{e}vy distribution in the central part followed by an
approximately exponential truncation\cite{YCX1}, and displays
power-law avalanche size distribution that agrees with the
real-life markets well\cite{YCX2}.

In succession, we investigate the scaling behavior of the stock
price time series generated by the present model. One can see
clearly in Fig. 3, the scaling of this artificial stock market is
$\delta_{model}=0.939\pm0.003$, which is excellently agree with
the real-life markets (see also the empirical data of DJIA for
comparison). This numerical results strongly suggest that the
present model has successfully reproduced some dynamical
characters of reality.

\section{Conclusion}
  In summary, by means of the DE method we investigate the scaling behavior
embedded in the time series of four typical financial markets. The
scale-invariance exponents are almost the same, being in the
interval of $[0.92, 0.95]$. This large deviation from the Gaussian
distribution reveals the strong correlations in the time series.
The present empirical study of the scaling behavior in real
markets also provides a usable quantity to check the reliability
of artificial market model.

Consequently, we propose a parsimonious percolation model for
stock market. Proper initial condition and parameters can lead its
stock price series being similar to the real-life index.
Especially, the scaling behavior detected with the DE method
agrees with the real-life financial markets very well.

\begin{acknowledgements}
This work has been supported by the National Science Foundation of
China under Grant No. 70271070, 70471033, 10472116, and the
Special Fund for the Doctoral Program of Higher Education under
Grant No. 20020358009.
\end{acknowledgements}


\begin{thebibliography}{Sornette2000}

\bibitem{Sornette2000} D. Sornette, Critical Phenomena in Natural Sciences, Berlin
Springer-Verlag, 2000.
\bibitem{Mandelbrot1963} B. B. Mandelbrot, J. Business 36(1963)
394.
\bibitem{Mantegna1995} R. N. Mantegna and H. E. Stanley, Nature 376(1995)
46.
\bibitem{Mantegna1997}  R. N. Mantegna and H. E. Stanley, Physica A 239(1997)
255.
\bibitem{Lo1999J} J. A. Lo and C. A. Mackinlay, Non-Random Walk Down Wall Street, Princeton
University Press, USA, 1999
\bibitem{Wang2001} B. H. Wang and P. M. Hui, Euro. Phys. J. B 20(2001)
573.
\bibitem{Stanley2001} H. E. Stanley, L. A. N. Amaral, X. Gabaix, P. Gopikrishnan, and V. Plerou, Physica A 299(2001)
1.
\bibitem{Ma1985} S. K. Ma, Statistic Mechanics, World
Scientific, Singapore, 1985.
\bibitem{Gnedenko1954} B. V. Gnedenko and A. N. Klomogorove, Limit Distributions for Sum of Independence
Random Variables, Addison Wesley, Reading, 1954.
\bibitem{Paladin1987} G. Paladin and A. Vulpiani, Physics Reports 156(1987) 147.
\bibitem{Peng1994} C. -K. Peng, S. V. Buldyrev, S. Havlin, M. Simons, H. E. Stanley, and A. L. Goldberger, Phys. Rev. E 49(1994) 1685
\bibitem{Scafetta2002} N. Scafetta and P. Grigolini, Phys. Rev. E 66(2002)
036130.
\bibitem{Ignaccolo20041} M. Ignaccolo, P. Allegrini, P. Grigolini, P. Hamilton, and B. J. West, Physica A
336(2004) 595.
\bibitem{Ignaccolo20042} M. Ignaccolo, P. Allegrini, P. Grigolini, P. Hamilton, and B. J. West, Physica A
336(2004) 623.
\bibitem{Grigolini2002} P. Grigolini, D. Leddon, and N. Scafetta, Phys. Rev. E 65(2002)
046203.
\bibitem{Bellazzini2003} J. Bellazzini, G. Menconi, M. Ignaccolo, G. Buresti, and P. Grigolini, Phys. Rev. E 68(2002) 026126.
\bibitem{Yang2004} H. -J. Yang, F. -C. Zhao, L. Qi, and B. -L. Hu, Phys. Rev. E 69 (2004) 066104.
\bibitem{Yang2005} H. -J. Yang, F. -C. Zhao, W. Zhang, and Z. -N. Li, Physica A 347(2005)
704.
\bibitem{Thaler1993} R. Thaler, Advances in Behavioral Finances,
Russell Sage Foundation, New York, 1993.
\bibitem{Lo1999} A. Lo, Financial Analysis J. 55 (1999) 13.
\bibitem{Friedman1991} D. Frideman, Econometrica 59 (1991) 637.
\bibitem{Zhou2005} T. Zhou, B. -H. Wang, P. -L. Zhou, C. -X. Yang,
and J. Liu, arXiv: cond-mat/0507626.
\bibitem{Wei2003} Y. -M. Wei, S. -J. Ying, Y. Fan and B. -H. Wang,
Physica A 325(2003) 507.
\bibitem{Zhou2004} T. Zhou, P. -L. Zhou, B. -H. Wang, Z. -N. Tang,
and J. Liu, Int. J. Mod. Phys. B 18 (2004) 2697.
\bibitem{Lettau1997} M. Lettau, J. Econ. Dyn.
Control 21 (1997) 1117.
\bibitem{Chen2001} S. -H. Chen, and C. -H. Yeh, J. Econ. Dyn.
Control 25 (2001) 363.
\bibitem{Grimmett1989} G. Grimmett, Percolation,
Springer-Verlag, Berlin, 1989.
\bibitem{Cont2000} R. Cont and J. P. Bouchaud, Macroeconomic Dynamics 4(2000)
170.
\bibitem{Stauffer2000} D. Stauffer, and N. Jan, Physica A 277
(2000) 215.
\bibitem{Castiglione2001} F. Castiglione, and D. Stauffer, Physica
A 300 (2001) 531.
\bibitem{Makowiec2004} D. Makowiec, P. Gnaci\'{n}ski, and M.
Miklaszewski, Physica A 331 (2004) 269
\bibitem{Wang2005} J. Wang, C. -X. Yang, P. -L. Zhou, Y. -D. Jin, T. Zhou, and B. -H. Wang, Physica A 354(2005)
505.
\bibitem{Eguiluz2000} V. M. Egu\'{i}luz and M. G. Zimmermann, Phys. Rev. Lett 85(2000)
5659.
\bibitem{Zhengdafang2002} D. -F. Zheng, G. J. Rodgers, P. M. Hui
and R. D'Hulst, Physica A 303(2002) 176.
\bibitem{zhengbo2004} B. Zheng, F. Ren, S. Tripmer and D. -F.
Zheng, Physica A 343(2004) 653.
\bibitem{Xie2005} Y. -B. Xie, B. -H. Wang, B. Hu, and T. Zhou,
Phys. Rev. E 71(2005) 046135.
\bibitem{Cavalcante2002} F. S. A. Cavalcante, A. A. Moreira, U. M. S. Costa, and J. S. Andrade Jr., Physica A 311(2002)
313.
\bibitem{Kahneman1979} D. Kahneman and A. Tversky, Econometrica 47(1979)
263.
\bibitem{Farmer2002} J. D. Farmer, Ind. Corp. Change 11(2002) 895.
\bibitem{YCX1} C. -X. Yang, J. Wang, T. Zhou, J. Liu, M. Xu, P.
-L. Zhou, and B. -H. Wang, Chin. Sci. Bull. (In Press).
\bibitem{YCX2} P. -L. Zhou, C. -X. Yang, T. Zhou, M. Xu, J. Liu,
and B. -H. Wang, New Mathematics and Natural Computation 1 (2005)
275.




\end{thebibliography}
\end{document}